\documentclass[twocolumn,showpacs,amsmath,nofootinbib,pra,aps,amssymb,longbibliography,superscriptaddress]{revtex4-2} 
\usepackage[T1]{fontenc}
\usepackage[utf8]{inputenc}
\usepackage{times}
\usepackage{color} 
\usepackage{array} 
\usepackage{amssymb,amsmath}
\usepackage{amsbsy}
\usepackage[pdftex]{graphicx} 
\usepackage{bm} 
\usepackage{float}
\usepackage{dcolumn}
\usepackage{booktabs}

\usepackage[unicode,breaklinks]{hyperref}
\hypersetup{
    unicode=true,
    plainpages=false, 
    colorlinks=true,
    linkcolor=blue,
    citecolor=blue,
    filecolor=black,
    urlcolor=blue
}
\usepackage{url}
\usepackage{verbatim}

\newcommand{\add}{a_{dd}}

\newcommand{\br}{\mathbf{r}}
\newcommand{\bq}{\mathbf{q}}

\newcommand{\bx}{\mathbf{x}}

\newcommand\gammaQF{\gamma_\mathrm{QF}}

\begin{document}
 
\title{Excitations and anisotropic sound in planar dipolar supersolids with tilted dipoles} 
 	\author{Reuben Cook} 
 	\author{Au-Chen Lee}  
 	\author{P. Blair Blakie} 
\affiliation{%
	Dodd-Walls Centre for Photonic and Quantum Technologies, Dunedin 9054, New Zealand}
\affiliation{Department of Physics, University of Otago, Dunedin 9016, New Zealand}
\date{\today} 
\begin{abstract}  
 We investigate the collective excitations of anisotropic dipolar supersolids in planar confinement, focusing on triangular and stripe phases in situations where the dipoles are titled to have a component in the plane. Using Bogoliubov–de Gennes calculations and  hydrodynamic theory, we identify the elastic parameters that govern the long-wavelength dynamics, including two orientational coefficients that capture the broken rotational symmetry induced by dipole tilt. Analytical expressions for the speeds of sound are obtained along the principal axes for triangular supersolids and along any propagation direction for the stripe supersolid. Our results provide a unified framework for understanding sound propagation in anisotropic dipolar supersolids and establish connections to recent experiments on sound propagation in striped Bose-Einstein condensates.
\end{abstract} 

\maketitle

\section{Introduction}
 
Dipolar Bose-Einstein condensates (BECs) are versatile systems for exploring the role of long-ranged dipole-dipole interactions (DDIs) on superfluids. In these systems the interactions arise from the magnetic moments of the atoms \cite{Griesmaier2005a,Mingwu2011a,Aikawa2012a,Miyazawa2022a} or from the electric dipole moments of the molecules \cite{Bigagli2024a} (also see \cite{Chomaz2022R}).
 One avenue of interest is how the anisotropic character of the DDIs manifest in the many-body physics. For example, revealing anisotropic effects in the dynamics \cite{Mishra2016a}, excitation spectrum \cite{Santos2003a, Blakie2013a,Bismut2012a,Baillie2015a}, superfluidity \cite{Ticknor2011a,Wenzel2018a}, solitons \cite{Tikhonenkov2008a}, vortices \cite{Yi2006a,Mulkerin2013a}, correlations and fluctuations \cite{Ticknor2012b,Baillie2014b,He2025a}.

Recently dipolar BECs have been used to produce supersolid states of matter \cite{Tanzi2019a,Bottcher2019a,Chomaz2019a}, in which crystalline order spontaneously emerges, and coexists with superfluid characteristics. Planar or pancake confined dipolar BECs, with the dipoles polarized along the tightly confined direction, have an interesting phase diagram with triangular, honeycomb and stripe types of crystalline order emerging dependent of the system parameters \cite{Zhang2019a,Norcia2021a,Poli2021a,Zhang2021a,Hertkorn2021b,Ripley2023a}. 
In such supersolid states new sound branches emerge from the spontaneously broken translational symmetries \cite{Watanabe2012a}. For example, the triangular and honeycomb supersolids exhibit three sound branches, including a transverse sound branch reflecting the shear modulus of the two-dimensional (2D) crystalline order \cite{Josserand2007a,Macri2013a,Watanabe2012a,Kunimi2012a,Poli2024b,Yapa2025a,Liebster2025a}. Here, because the dipole is normal to the plane, the effect of the dipole interactions are invariant to planar rotations. As a result the long wavelength properties of the 2D supersolid states are isotropic. This is not the case for the stripe state, which spontaneously breaks the planar rotational symmetry.

In addition to the planar regime, where  the confinement dominates along one direction, but the system remains three-dimensional, there has been considerable theoretical work on the tightly confined limit, where the dipoles are restricted to move in a plane. This pure-2D regime, has been simulated with Monte Carlo methods, which have explored issues around Berezinskii-Kosterlitz Thouless transitions and the formation of stripes \cite{Macia2012a,Fedorov2014a,Bombin2017a,Bombin2019a,Cinti2019a,Staudinger2023a}. A key ingredient in favour the formation of stripes is tilting the dipoles to have a component in the plane. This introduces an anisotropy in the plane, leading to a preferred orientation for the stripes.

The use of dipole tilt was also explored theoretically and experimentally in the cross-over from the cigar-shaped to pancake-shaped (planar) harmonic trap in Ref.~\cite{Wenzel2017a}, where it was demonstrated that the tilt could be used to prepare incoherent stripe states. More recently attention has focused the application of dipolar tilt on the production of supersolids in the planar regime relevant to current experiments \cite{Lima2025a,Sanchez-Baena2025a} (also see \cite{Aleksandrova2024a}). Notably, Lima \textit{et al.}~\cite{Lima2025a} used an approximate supersolid wave function to characterize the phase diagram, showing that dipole tilt shifts the phase boundaries between the various 2D phases (i.e.,~the uniform superfluid, stripe, triangular and honeycomb phases) compared to the untilted case \cite{Zhang2019a,Ripley2023a}. It was also shown that it can significantly broaden the parameter range over which continuous transitions occur between the phases. Such continuous transitions are experimentally favourable, because they allow parameter ramps (e.g., of the scattering length, dipole tilt, or trap confinement) to smoothly take the system between different phases.

Beyond ground state properties \cite{Aleksandrova2024a,Lima2025a,Sanchez-Baena2025a}, little is understood about  planar supersolid states with dipolar tilt. In this work we address this by examining their excitation spectrum, focussing on how anisotropy arising controlled by  dipole tilt manifests in the speeds of sound. We present results for triangular and stripe supersolid phases, and we compare of the excitations spectrum obtained by direct numerical calculations of the Bogoliubov-de Genes (BdG) equations to an anisotropic hydrodynamic theory that we develop. This supersolid hydrodynamic theory \cite{Andreev1969a,Son2005a,Josserand2007b,Yoo2010a} (also see \cite{Hofmann2021a}) has additional elastic parameters compared to the isotropic theories that have been used previously (e.g., see \cite{Heinonen2019a,Buhler2023a,Platt2024a,Poli2024b,Rakic2024a,Sindik2024a,Blakie2025a}), including two new orientational parameters to describe the broken rotational symmetry from the dipole tilt. Here we focus on using the hydrodynamic theory to predict the speeds of sound, however it can also be used to understand the long-wavelength fluctuations of the supersolid \cite{Yoo2010a,Platt2024a}.

In general the speeds of sound have a complicated dependence on the elastic parameters, but we obtain analytic results for the speeds of sound along the principal axes for the 2D crystalline cases, and in any direction for the stripe case. The stripe results also allow us to make connection to recent experiments where a (non-supersolid) stripe state was produced using a optical lattice, for the purpose of determining the superfluid fraction \cite{Chauveau2023a,Tao2023a}. In these systems there is only a single sound branch and a key observation was the anisotropic speed of sound, which was used to reveal the anisotropic superfluid response. We can connect our anisotropic supersolid hydrodynamic theory to these results by considering the limit that the crystal becomes rigid.

The outline of the paper is as follows.
In Sec.~\ref{Sec:SysExc} we introduce the planar dipolar system with tilt and show examples of a supersolid state and its excitation spectrum. Details of the extended mean-field theory used for the ground state and excitation calculations is given in Appendix \ref{Sec:EMFthry}.
Then in Sec.~\ref{Sec:HydroFormalism} we introduce the relevant hydrodynamic fields for the planar supersolids states, and use these to define the elastic parameters. This allows us to introduce a quadratic supersolid Lagrangian which we use to obtain equations of motion for the hydrodynamic fields.  Considering plane wave solutions in Sec.~\ref{Sec:Sound} we obtain general equations for the speeds of sound. We then specialise these to obtain analytic results for supersolids with 2D and 1D crystalline structure.  In Sec.~\ref{Sec:Results} we present results comparing direct numerical calculations of the speeds of sound from the BdG spectrum to the results of the hydrodynamic theory. We also observe the typical behavior of the elastic parameters in dipolar supersolids and how they tend to change with dipolar tilt.
We then conclude in Sec.~\ref{Sec:concl}.

\section{System and excitations}\label{Sec:SysExc}
 We consider a dipolar BEC axially confined by a harmonic trap potential of frequency $\omega_z$.  The atoms are free to move in the $xy$-plane, and are constrained to have an average areal density of $\rho_0$.  The atoms interact with a short-range contact interaction, characterized by the $s$-wave scattering length $a_s$. An external magnetic field is used to polarize the magnetic moment of the atoms in a direction lying in $xz$-plane, at an angle of $\alpha$ to the $z$ axis.  The resulting DDIs between the atoms is characterized by the length scale $a_{dd}$, with a value of $130.8\,a_0$  for $^{164}$Dy, where $a_0$ is the Bohr radius.

  \begin{figure}[htbp] 
    \centering
    \includegraphics[width=3.4in]{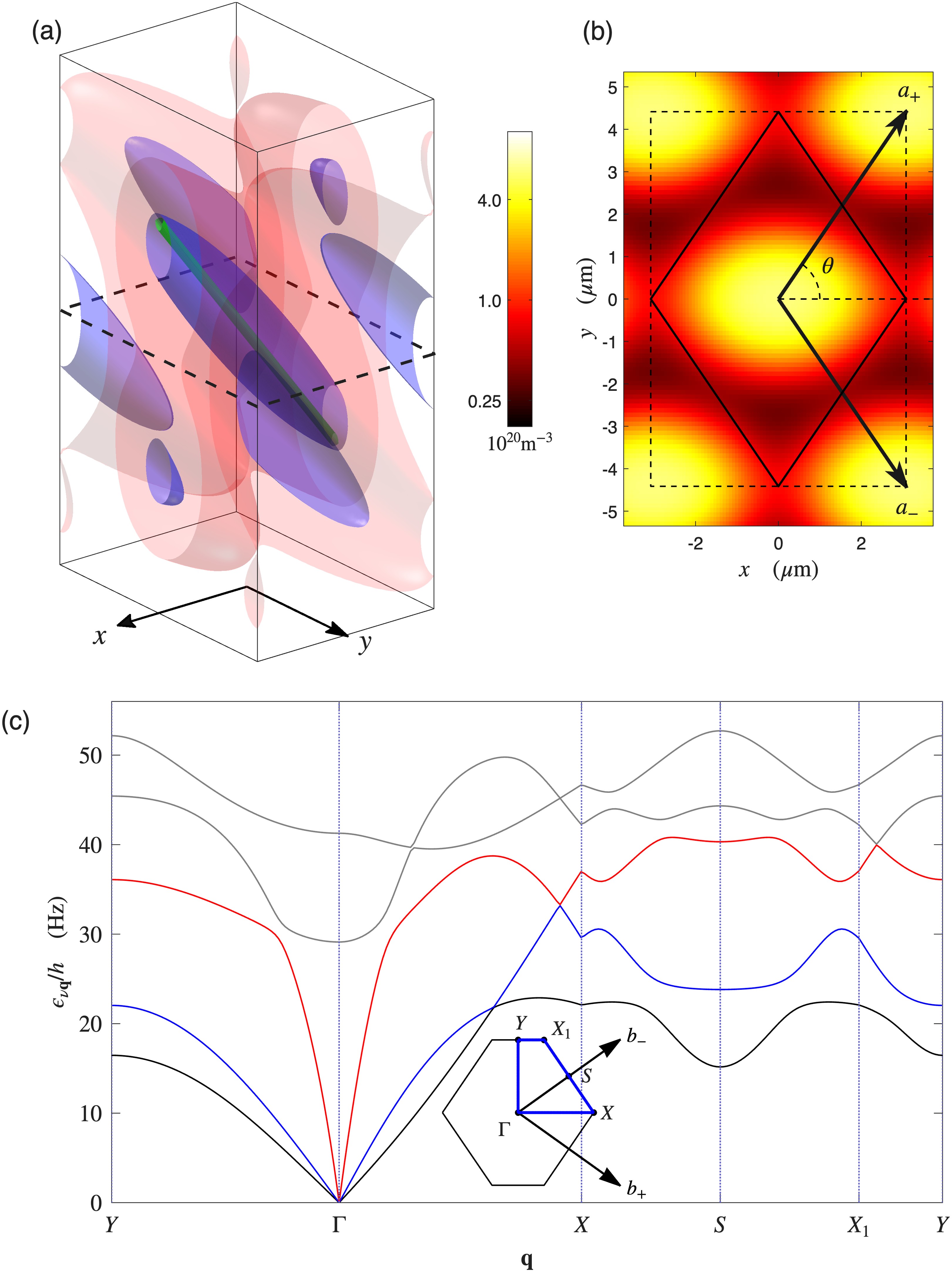} 
    \caption{Ground state and band structure of an anisotropic 2D dipolar supersolid supersolid.  Ground state  (a) density isosurface and (b) $z=0$ density slice. (c) The band structure along the special lines, with the first Brillouin zone and special points shown for reference.  In (a) the green line indicates the direction of the dipole polarization. The red (blue) isosurface are at $0.5\times10^{20}/$m$^{3}$ ($4\times10^{20}/$m$^{3}$). The dashed line indicates the conventional cell.  In (b) the primitive cell (solid line), conventional cell (dashed line) and the lattice vectors are shown. 
     Parameters: $^{164}$Dy gas with $a_{dd}=130.8\,a_0$, $a_s=100.7\,a_0$, $\rho_0=0.075/a_{dd}^2$ and axial confinement of $\omega_z/2\pi=72.4\,$Hz and the dipoles tilted by an angle of $\alpha=38^\circ$. The lattice vector angle is $\theta=55.1^\circ$ with length $a=5.38\,\mu$m.} 
    \label{fig:gsbs}
 \end{figure}

The system is analyzed using an extended mean-field theory that takes into account the beyond-mean-field corrections.  The  condensate wave function $\Psi$ is determined by minimising the energy density functional $\mathcal{E}$. In regimes where the condensate wave function exhibits crystalline structure, it is calculated in a unit cell with lattice vectors $\{\mathbf{a}_\beta\}$ reflecting the periodicity. The phase diagram for this system has been considered for untilted dipoles ($\alpha=0$) in Refs.~\cite{Zhang2019a,Ripley2023a}, and in for the tilted case in Ref.~\cite{Lima2025a}. The phase diagrams show that a modulated state emerges when  $a_s$ is reduced to a value sufficiently far below $a_{dd}$. The nature of the modulated state depends on the value of $a_s$ and $\rho_0$ and can exhibit either a 1D (stripe state) or a 2D (triangular-like or honeycomb-like) lattice structure.  The excitations of the supersolid states are determined by solving the BdG equations. Numerically solving these equations yields the excitation energies $\epsilon_{\nu\mathbf{q}}$ with $\nu$  being a band index and $\mathbf{q}=(q_x,q_y)$ being a planar quasimomentum. As the extended mean-field theory for the stationary states, excitations and numerical calculations have been discussed elsewhere (see \cite{Poli2024b,Blakie2025a}), these are summarized in the Appendix \ref{Sec:EMFthry}.

We show an example of 2D supersolid for tilted dipoles in Fig.~\ref{fig:gsbs} and give its band structure. Here the effect of dipole tilt is to distort the unit cell from being triangular, to being orthotropic [see Figs.~\ref{fig:gsbs}(a) and (b)]. This generally causes the angle $\theta$ of the lattice vectors to decrease to a value of less than $60^\circ$ [see Fig.~\ref{fig:gsbs}(b)]. Associated anisotropy emerges in the excitation spectrum, with the speeds of sound along the principal axes being different, i.e.~the slopes of the gapless excitation branches along $q_x$ (i.e.,~$\Gamma$-$X$) are different to those along $q_y$ (i.e.,~$\Gamma$-$Y$) in Fig.~\ref{fig:gsbs}(c). We later analyse these differences in detail [see Sec.~\ref{Sec:anisotriresults} and  Fig.~\ref{fig:tri}].

\section{Hydrodynamic Formalism}  \label{Sec:HydroFormalism}
\subsection{Hydrodynamic fields}
The hydrodynamic description of a system involves a number of  ``slow variables" that govern its long-wavelength behavior. These are typically fields associated with conservation laws and broken symmetries. For a supersolid we require three distinct fields for this purpose. The first two fields are the areal density $\rho(\bm{\rho})=\rho_0 + \delta\rho(\bm{\rho})$, and the superfluid phase $\phi(\bm{\rho})$, where  $\bm{\rho}=(x,y)$ is the planar coordinate. These are associated with the conservation of number and the broken U(1)-gauge symmetry, and occur in the description of ordinary superfluids. The broken translational invariance arising from the crystalline order introduces the displacement  $\mathbf{u}(\bm{\rho})=(u_x,u_y)$ as the third field. This describes the spatial displacement of the crystal from its equilibrium location. These fields correspond to coarse-grained quantities obtained from the microscopic description of the system (e.g.,~see discussion in Ref.~\cite{Platt2024a}). For the stripe supersolid, translational symmetry is only broken in one spatial dimension, and the displacement field reduces to being a single component field, which we take to be $u_y$ (see Sec.~\ref{Sec:1DLatticeCell}).

\subsection{Elastic parameters}\label{Sec:ElasticParams}
The elastic parameters describe the response of the supersolid to changes in the hydrodynamic variables. 
These parameters appear in the Lagrangian description of the supersolid and allow us to characterise its excitations. 
We determine the elastic parameters from our microscopic (extended mean-field theory) calculations of the ground states.
In the long-wavelength limit, changes in the three fields correspond to distorting the unit cell, twisting the phase of the wave function boundary conditions or changing the average density constraint. 
Indeed, we consider the energy density as function of these changes $\mathcal{E}(\rho_0+\delta\rho,\mathbf{v},\bm{\nabla}_{\bm{\rho}}\mathbf{u})$, with the ground state value being $\mathcal{E}_0=\mathcal{E}(\rho_0,0,0)$. Note that system energy is independent of a global phase change or a uniform spatial displacement, and thus only depends on gradients of these fields, i.e., the superfluid velocity $\mathbf{v}=\frac{\hbar}{m}\bm{\nabla}_{\bm{\rho}}\phi$, and the strain $\bm{\nabla}_{\bm{\rho}}\mathbf{u}$. 

 Below we identify the relevant elastic parameters and discuss how they can be determined from calculations of stationary states. We first focus on the 2D crystalline case before considering the 1D case relevant to the stripe state.

\subsubsection{2D Lattice: elastic tensor and orientational terms}\label{Sec:LatticeCell}
\begin{figure}[htbp] 
    \centering
    \includegraphics[width=2.85in]{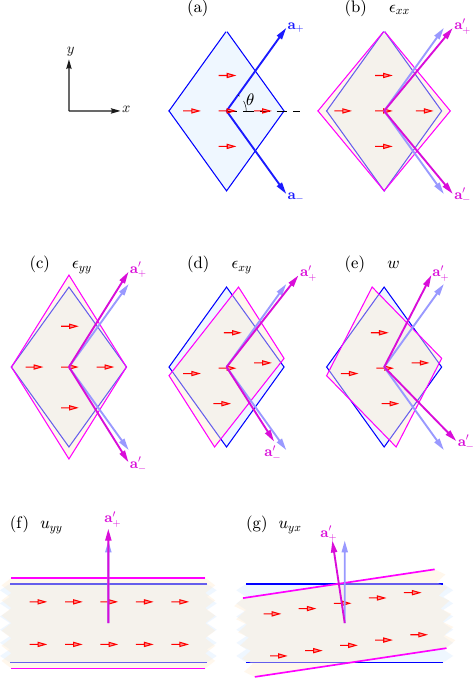} 
\caption{Primitive cells and distortions. (a) Primitive 2D unit cell used in subplots (b)-(e) to give examples of: (b) Normal strain in the $x$-direction; (c) Normal strain in the $y$-direction; (d) Shear strain; (e) Rotation. The 1D stripe cell is illustrated subject to: (f) Normal strain in the $y$-direction; (g) Rotation. In (b)-(g) the blue cells and grey vectors illustrate the unperturbed cell and lattice vectors, while the magenta cells and vectors represent the perturbed cell and lattice vectors. The horizontal red vectors illustrate the component of the tilted dipoles in the $xy$ plane. }
\label{fig:cell}
\end{figure}

In general, the ground state unit cell is orthorhombic (centered rectangular), with two orthogonal symmetry axes (principal axes). Because the tilted dipoles have a component along $x$, the $x$- and $y$-axes are the principal axes. The direct lattice vectors for the unit cell are
 \begin{align}
\mathbf{a}_{\pm}&=a(\cos\theta\hat{\mathbf{x}}\pm\sin\theta\hat{\mathbf{y}}),\label{apm}
\end{align}
e.g.,~see Figs.~\ref{fig:gsbs}(b) and \ref{fig:cell}(a). The area of the unit cell is $A_{\mathrm{uc}}=a^{2}\sin2\theta$, and the  reciprocal lattice vectors  are 
\begin{align}
\mathbf{b}_{\pm}=\frac{2\pi\csc2\theta}{a}(\sin\theta\hat{\mathbf{q}}_x\mp\cos\theta\hat{\mathbf{q}}_y),
\end{align} 
see Fig.~\ref{fig:gsbs}(c).
For  $\alpha=0$ the energy minimising unit cell for the 2D crystalline state has $\theta=60^\circ$. This case has hexagonal symmetry and the various elastic properties become isotropic (see Sec.~\ref{Sec:Iso}).

We solve for the ground states on a primitive unit cell specified by the relevant lattice vectors $\{\mathbf{a}_\beta\}$. Straining the unit cell changes the spatial coordinates according to $\rho^\prime_i =\rho_i + u_{ij}\rho_j$. Here the indices are $i,j\in\{x,y\}$, $u_{ij}=\partial_ju_i$   is the strain tensor field, and we assume the Einstein summation convention. It is useful to introduce the symmetric strain tensor
$\epsilon_{ij}=\frac{1}{2}(u_{ij}+u_{ji})$, and the corresponding antisymmetric tensor, which for our 2D case is the single quantity $w=-\frac{1}{2}(u_{xy}-u_{yx})$  describing an anti-clockwise rotation in the plane by angle $w$. Thus, deformations of the unit cell for the 2D supersolid are parameterised by $\{\epsilon_{xx},\epsilon_{yy},\epsilon_{xy},w\}$, which distort the lattice vectors according to the linear transformation
\begin{align}
\mathbf{a}_\beta^\prime=\begin{bmatrix}
1+\epsilon_{xx} & \epsilon_{xy}-w \\
 \epsilon_{xy}+w & 1+\epsilon_{yy}
\end{bmatrix}\mathbf{a}_\beta.\label{Alatt}
\end{align} 
For reference, in Figs.~\ref{fig:cell}(b)-(f) we show pure deformation of each kind on the unit cell.
These demonstrate the kinds of small changes we make to the unit cell to determine the elastic parameters using finite difference derivatives of the stationary state energy density. 

The lattice elastic tensor $C_{ijkl}$ is obtained as the second derivative respect to the strain \cite{Bavaud1986a}. 
 This tensor has the symmetries $C_{ijkl}=C_{jikl}=C_{ijlk}=C_{klij}$, and the 2D orthorhombic cell has four non-zero independent parameters $\{C_{xxxx},C_{yyyy},C_{xxyy},C_{xyxy}\}$. 
This allows the elastic energy to be described by the quadratic form
\begin{align}
\mathcal{E}_\mathrm{el}=&\frac{1}{2}C_{xxxx}\epsilon_{xx}^2+\frac{1}{2}C_{yyyy}\epsilon_{yy}^2 +2C_{xyxy}\epsilon_{xy}^2\label{Eel2D}\\
&+C_{xxyy}\epsilon_{xx}\epsilon_{yy}.\nonumber
\end{align} 
 
 When the dipole moment has a non-zero component in the plane, the planar rotational symmetry is broken and the system energy will depend on $w$. This introduces an orientational energy
 \begin{align}
 \mathcal{E}_\mathrm{ori}=\frac{1}{2}Aw^2+B\omega\epsilon_{xy}\label{Eori2d}
 \end{align}
where
 \begin{align}
 A=\frac{\partial ^2\mathcal{E}}{\partial w^2},\label{Arot}
 \end{align}
is an orientational constant, and
 \begin{align}
 B=\frac{\partial ^2\mathcal{E}}{\partial w\partial\epsilon_{xy}},\label{Brot}
 \end{align} 
describes the coupled effect of shear strain and rotation.

\subsubsection{1D Lattice: elastic tensor and orientational terms}\label{Sec:1DLatticeCell}
We also examine stripe states with crystalline structure defined by a single lattice vector. For $\alpha=0$ the system is invariant to rotations in the plane, and the stripes can be oriented in any direction. However, when the dipoles have a component along $x$ it is energetically favourable for the stripes to orient along $x$, i.e.,~the single lattice vector will be $\mathbf{a}_+=a\hat{\mathbf{y}}$ [e.g.,~see Fig.~\ref{fig:cell}(f)].

Deformations of the stripe state are described by a single displacement field $u_{y}$, and thus the energy will only depend the two derivatives $u_{yy}$ and $u_{yx}$. The first is a normal strain along $y$, i.e.,~$\epsilon_{yy}=u_{yy}$, whereas the second can be interpreted as a rotation $w=u_{yx}$.  
Thus, deformations of the unit cell for the stripe supersolid are parameterised by $\{\epsilon_{yy},w\}$, which distort the lattice vector according to the linear transformation
\begin{align}
\mathbf{a}_+^\prime=\begin{bmatrix}
1  &  -w \\
 w & 1+\epsilon_{yy}
\end{bmatrix}\mathbf{a}_+.\label{Alatt1D}
\end{align} 
Examples for both types of distortion are shown in Figs.~\ref{fig:cell}(g)-(h).
The elastic and orientational terms are simpler than for the 2D lattice [cf.~Eqs.~(\ref{Eel2D}) and (\ref{Eori2d})], reducing to
\begin{align}
\mathcal{E}_\mathrm{el}=\frac{1}{2}C_{yyyy}\epsilon_{yy}^2, \qquad 
 \mathcal{E}_\mathrm{ori}=\frac{1}{2}Aw^2.\label{Eel1D}
 \end{align}

\subsubsection{Density: bulk modulus and density-strain}
By varying the average system density, we can evaluate the bulk modulus
\begin{align}
\alpha_{\rho\rho}= \frac{\partial ^2\mathcal{E}}{\partial \rho^2}.
\label{alpharr}
\end{align}
This can be equivalently evaluated as $\alpha_{\rho\rho}=\partial\mu/\partial\rho$, where $\mu$ is the chemical potential.
The bulk modulus relates to the fluid compressibility at constant strain as $\tilde\kappa=(\rho_0^2\alpha_{\rho\rho})^{-1}$.
We also consider a mixed derivative of density and strain
 \begin{equation}
   \gamma_{ij}=\frac{\partial^2 \mathcal{E}}{\partial \epsilon_{ij} \partial \rho}.
\end{equation}
In general, these density-strain terms are small and often neglected (e.g.,~see \cite{Hofmann2021a}), although they are important in obtaining quantitative agreement between hydrodynamic theory and BdG predictions (see \cite{Platt2024a}). For the 2D supersolid, we find that  $\gamma_{ij}$ is only non-zero for normal strains i.e.,~$\{\gamma_{xx},\gamma_{yy}\}$,  reflecting the coupling between change in cell volume and density.  For the stripe case we only have the single parameter $\gamma_{yy}$. 

\subsubsection{Phase: superfluid fraction}
The emergence of a density modulation (i.e.,~broken translational symmetry) reduces the superfluid fraction of a BEC from unity at zero temperature \cite{Leggett1970a,Leggett1998a,Sepulveda2010a,Blakie2024a}. This feature has been recently explored in experiments with stripe BECs \cite{Chauveau2023a,Tao2023a} and dipolar supersolids \cite{Biagioni2024a}, and more recently in 2D spatially modulated BECs \cite{Rabec2025a}.
 The superfluid fraction can be determined by the energetic response of the system to a twist in the phase (see Ref.~\cite{Saslow1976a,Sepulveda2010a,Blakie2024a,Fisher1973a}), and for a 2D system it is generally described by a second-order tensor.
This can be specified by twisting the phase of the boundary conditions $\Psi(\bm{\rho}+\mathbf{a}_\beta,z)=\Psi(\bm{\rho},z)e^{i\Delta\phi_\beta}$, where $\Delta\phi_\beta=m\mathbf{v}\cdot\mathbf{a}_\beta/\hbar$, and $\mathbf{v}$ is the (planar) superfluid velocity.  In general, the superfluid density tensor is given by
\begin{align}
\rho_{s,ij}=\frac{1}{m}\frac{\partial^2\mathcal{E}}{\partial v_i\partial v_j},
\end{align}  
and the associated normal density tensor is 
\begin{align}
\rho_{n,ij}=\rho_0\delta_{ij}-\rho_{s,ij}.
\end{align}

\subsection{Supersolid Lagrangian}
The quadratic supersolid Lagrangian describing the linearized evolution of the  hydrodynamic fields is  
\begin{align}
\mathcal{L} 
=&-\hbar\delta\rho\partial_{t}\phi-\rho_0\frac{\hbar^{2}}{2m} (\partial_{i}\phi)^{2}-\delta\rho\sum^\prime_{ij}\gamma_{ij}\epsilon_{ij}-\frac{1}{2}\alpha_{\rho\rho}\delta\rho^{2}  \nonumber\\  
 & +\frac{1}{2}m\sum^\prime_{ij}\rho_{n,ij}\left(\partial_{t}u_{i }-\frac{\hbar}{m}\partial_{i}\phi\right)\left(\partial_{t}u_{j}-\frac{\hbar}{m}\partial_{j}\phi\right) \nonumber\\ 
 &
 -\mathcal{E}_\mathrm{el}-\mathcal{E}_\mathrm{ori}\label{eq:FullLagrangian} 
 \end{align} 
where $\partial_t$ denotes the time derivative. The primed summation indicates that the indices are summed over the components of the displacement field (i.e., $\{x,y\}$ for the 2D supersolid, and $\{y\}$ for the stripe supersolid), for other terms we employ the Einstein summation convention with summation taken over both planar components. 
This Lagrangian is of the form developed by Yoo and Dorsey in Ref.~\cite{Yoo2010a} (also see \cite{Son2005a}), but augmented by the orientational terms we have introduced to describe the effect of a planar component of the dipoles. 

Since no derivatives of $\delta\rho$ appear in Eq.~(\ref{eq:FullLagrangian}), so we can use its Euler-Lagrange equation, 
\begin{align}
\delta\rho=-(\hbar\partial_{t}\phi+\sum^\prime_{ij}\gamma_{ij}\epsilon_{ij})/\alpha_{\rho\rho},
\end{align}
 to eliminate $\delta\rho$ and yield the effective Lagrangian: 
\begin{align}
\mathcal{L}  &=\frac{\hbar^{2}}{2\alpha_{\rho\rho}}(\partial_{t}\phi)^{2}-\rho_{s,ij}\frac{\hbar^{2}}{2m}\partial_{i}\phi\partial_{j}\phi+\frac{1}{2}m\sum^\prime_{ij}\rho_{n,ij}\partial_{t}u_{i}\partial_{t}u_{j}\nonumber\\
&\, -\tilde{\mathcal{E}}_\mathrm{el}-\mathcal{E}_\mathrm{ori}-\hbar\sum^\prime_{ij}\tilde{\rho}_{ij}\partial_{t}u_{i}\partial_{j}\phi,   \label{EqredL}
\end{align}
where we introduced the renormalized parameters. 
\begin{align}
\tilde{C}_{ijkl} & =C_{ijkl}-\frac{\gamma_{ij}\gamma_{kl}}{\alpha_{\rho\rho}},\quad
\tilde{\rho}_{ij}  =\rho_{n,ij }-\frac{\gamma_{ij}}{\alpha_{\rho\rho}},
\end{align}
with $\tilde{\mathcal{E}}_\mathrm{el}$ corresponding to $\mathcal{E}_\mathrm{el}$ Eq.~(\ref{Eel2D}) and (\ref{Eel1D}) for the 2D and stripe supersolids, respectively, but with $C_{ijkl}$ replaced by $\tilde{C}_{ijkl}$.
From Eq.~(\ref{EqredL}) the Euler-Lagrange equations are 
\begin{align}
	 {\hbar^{2}}\left(\frac{1}{\alpha_{\rho\rho}}\partial_{t}^{2}-\frac{\rho_{s,ij}}{m}\partial_{i}\partial_{j}\right)\phi=\hbar\sum^\prime_{ij}\tilde{\rho}_{i  j }\partial_{t}\partial_{i}u_{j}	,\label{EL1}\\  
\sum^\prime_{k}\left(m\rho_{n,i k  }\partial_{t}^{2}-[\tilde{C}_{i j k l}  +\Omega_{i  jk  l}]\partial_{j}\partial_{l}\right)u_{k }
=\hbar\sum^\prime_{ij}\tilde{\rho}_{ij}\partial_{t}\partial_{j}\phi,\label{EL2}
\end{align}
where  
\begin{align} 
\Omega_{ijkl}&=\frac{1}{4}[A- B(\epsilon_{ijz}+\epsilon_{klz})]\epsilon_{ijz}\epsilon_{klz},&  \mbox{2D}\\
 \Omega_{yjyl}&= A\delta_{jx}\delta_{lx},&\,\mbox{stripe}\label{OmegaStripe}
\end{align} 
with $\epsilon_{ijz}$ being the Levi-Civita tensor. The expanded form of the Euler-Lagrange equations for both cases are given in Appendix \ref{Sec:AppendEL}.

The homogeneous version of Eq.~(\ref{EL1}) is the usual superfluid hydrodynamic equation, whereas the homogeneous version of Eq.~(\ref{EL2}) is the Navier-Cauchy equation describing the evolution of the displacement field for a linear elastic material.  The right hand sides of these equations couple the superfluid and elastic degrees of freedom via the normal component. 

\section{Speeds of sound}\label{Sec:Sound}

Having introduced the system, elastic parameters and hydrodynamic equations, we are now in a position to address the speeds of sound. We can determine these by seeking plane-wave solutions of the hydrodynamic equations of motion of the form 
\begin{align}
\mathbf{Q}(\bm{\rho},t)\equiv
\begin{bmatrix}
\phi(\bm{\rho},t) \\
\mathbf{u}(\bm{\rho},t )
\end{bmatrix}
=\mathbf{Q}_{\nu}e^{i(\mathbf{q}\cdot\bm{\rho}-\omega t)},
\end{align}
 where $\mathbf{Q}_{\nu}$ is a constant vector.  This describes coupled phase and displacement waves propagating through the supersolid.
 Substituting this ansatz into the Euler-Lagrange equations (\ref{EL1}) and (\ref{EL2}) yields the system of equations
  $M\mathbf{Q}_\nu=0$, where  
\begin{align}
M&=\begin{bmatrix}
M_{\phi\phi} & M_{\phi \mathbf{u}} \\
M_{\mathbf{u}\phi} &M_{\mathbf{u} \mathbf{u}}
\end{bmatrix},\label{Mgen}\\
M_{\phi\phi}&= {\hbar^{2}} \left(\frac{\rho_{s,xx}q_{x}^{2}}{m}+\frac{\rho_{s,yy}q_{y}^{2}}{m}-\frac{\omega^{2}}{\alpha_{\rho\rho}}\right),\\  
M_{\phi u_{i }}&=-\hbar\omega q_{i }\tilde{\rho}_{i  i},  \\ 
M_{u_i  u_k }
&=(\tilde{C}_{ijkl} +\Omega_{ijkl})q_j  q_l 
	-	m\,\omega^2\,\rho_{n,ik}.
\end{align}
A non-trivial plane wave solution requires that the matrix $M$ is singular. This is satisfied for particular ratios of the frequency and wave vector, $\omega=c_\nu q$, obtained by solving $\mathrm{det}M =0$. This is of the form of a quadratic eigenvalue problem \cite{TisseurMeerbergen2001}, and its solution yields up to 3 distinct speeds of sound $c_\nu$. For anisotropic systems, these speeds of sound can depend on direction, i.e.,~$c_{\nu}\left(\hat{\mathbf{q}}\right)$. Obtaining the speeds of sound in the general case requires solving a cubic equation and is best done numerically.  However, we can identify several special cases that we can solve analytically, because $M$ simplifies to a block diagonal form.

We now specialise this result to several different cases. For context, we start with the isotropic case, which has been considered previously \cite{Josserand2007b,Yoo2010a,Rakic2024a,Poli2024b,Blakie2025a}, before presenting the anisotropic results we develop here.
 
\subsection{Isotropic 2D supersolid}\label{Sec:Iso}
 A supersolid of untilted dipoles in the triangular or honeycomb phase is intrinsically isotropic. Here the superfluid fraction, elastic tensor, and density-strain are isotropic, i.e.~
 \begin{align}
 \rho_{s,ij}&=\rho_s\delta_{ij},\\
 C_{ijkl}&= \lambda\delta_{ij}\delta_{kl}+ \mu(\delta_{ik}\delta_{jl}+\delta_{il}\delta_{jk}),\label{Ciso}\\
    \gamma_{ij}&=\gamma\delta_{ij},
 \end{align}
 where $\{\lambda,\mu\}$ are the Lam\'e parameters \cite{LandauElasticity}.  As a result $\rho_{n,ij}=(\rho_0-\rho_s)\delta_{ij}$ and the renormalized parameters $\tilde{C}_{ijkl}$ and $\tilde{\rho}_{ij}$ are also isotropic, notably $\tilde{C}_{ijkl}$ takes the  same form as (\ref{Ciso}), but with a renormalized  Lam\'e parameter $\tilde{\lambda}=\lambda-\gamma^2/\alpha_{\rho\rho}$.
 In this situation $M_{\mathbf{u}\mathbf{u}}$  simplifies to:
 \begin{align}
 M_{\mathbf{u}\mathbf{u}}=(\mu+\tilde{\lambda})\hat{\mathbf{q}}^T_l\hat{\mathbf{q}}_l+(\mu q^2-m\omega^2\rho_{n})I_2,
 \end{align}
 where $I_2$ is the 2$\times$2 identity matrix, and $\hat{\mathbf{q}}_l=\mathbf{q}/q$ is the longitudinal unit propagation vector.
Decomposing the displacement field into longitudinal and transverse components as $\mathbf{u}= u_l\hat{\mathbf{q}}_l+u_t\hat{\mathbf{q}}_t$, where $\hat{\mathbf{q}}_t$ are the planar unit vectors  perpendicular to $\mathbf{q}$, the full $M$ matrix (\ref{Mgen}) becomes block diagonal reducing to the two uncoupled problems:
 \begin{align} 
 M_l&=\!\begin{bmatrix}\hbar^{2}\left(
 \frac{\rho_s q^2}{m} -\frac{\omega^2}{\alpha_{\rho\rho}}\right) &
  -\hbar\omega q\tilde{\rho}  \\
   -\hbar\omega q\tilde{\rho} & 
  (\tilde\lambda+2\mu)q^2-m\omega^2\rho_n \label{Mliso}
 \end{bmatrix},\\
 M_t&=\mu q^2-m\omega^2\rho_{n}.\label{Mtiso}
 \end{align}
Solving $\det M_l=0$ determines the longitudinal speeds of sound, while solving $M_t=0$ determines the transverse speed of sound.  

This general structure we have arrived at repeats for the other analytic results we present in this paper, it is useful to define a standard solution. The diagonal matrix elements of $M_l$ are proportional to $c_\alpha^2q-\omega^2$, where $c_\alpha$ are the two (uncoupled) speeds of sound
  \begin{align}
 c_s^2 &=\rho_s\alpha_{\rho\rho}/m,\\
 c_l^2 &=(\tilde{\lambda}+2\mu)/m\rho_n.
 \end{align}
Here $ c_s$ is the speed of sound for a uniform superfluid of superfluid density $\rho_s$ and $c_l$ is the longitudinal elastic speed of sound for a crystal of density $\rho_n$. This allows us to write the singular condition for $M_l$ as
 \begin{align}
 \det M_l\propto (c_s^2-c^2)(c_l^2-c^2)-c^2g^2=0,\label{EgMl}
 \end{align}
where  $g^2= \tilde{\rho}^2\alpha_{\rho\rho}/m\rho_{n}$ is the superfluid-lattice coupling and we have set $c=\omega/ q$ to determine the coupled speeds of sound. Solving Eq.~(\ref{EgMl}) yields the two solutions for $c$:
 \begin{align}
c_\pm^2=\frac{1}{2}\left[c_s^2+c_l^2+g^2\pm\sqrt{(c_s^2+c_l^2+g^2)^2-4c_s^2c_l^2}\right].\label{gencpm}
\end{align}
In the absence of coupling, $g\to0$, this yields the trivial result of uncoupled superfluid and crystal sound waves $c_\pm\to\{c_s,c_l\}$. In general, the longitudinal waves have coupled crystal and superfluid dynamics.  Note that because the density-strain coupling is typically small, we can approximate $g^2\approx\rho_n\alpha_{\rho\rho}/m$, and observe that for $\rho_n\sim\rho_s$, this coupling term is of similar magnitude to $c_s$, so the effect of coupling is usually very significant.
We also mention that the longitudinal sound waves in a 1D and isotropic 2D supersolids have the same basic behavior, and is discussed further in Refs.~\cite{Platt2024a,Poli2024b}. 

The singular condition for the transverse displacement equation (\ref{Mtiso}) can similarly be written as
\begin{align}
M_t\propto c_t^2-c^2=0,
\end{align}
with $c_t^2=\mu/\rho_n$ being the transverse elastic speed of sound for a crystal of density $\rho_n$.

\subsection{Anisotropic 2D supersolid}
The excitations for the general 2D anisotropic supersolid are described by the matrix introduced in Eq.~(\ref{Mgen}). In general, the determination of sound waves requires a numerical solution of a quadratic eigenvalue problem. \cite{TisseurMeerbergen2001}. However, for propagation along the principal axes, the sound waves are longitudinal or transverse, and the problem reduces to a similar form to the isotropic case.

\subsubsection{Propagation along the $x$-principal axis}
We solve for the sound waves propagating along $\Gamma$-$X$ with $\hat{\mathbf{q}}_l=\hat{\mathbf{q}}_x$. Here (\ref{Mgen}) simplifies to the two decoupled blocks
\begin{align} 
M_l&=\begin{bmatrix}\hbar^{2}\!\left(\!
\frac{\rho_{s,xx}q^{2}}{m}-\frac{\omega^{2}}{\alpha_{\rho\rho}}\!\right)    & 
-\hbar\omega q\tilde{\rho}_{xx} \\
-\hbar\omega q\tilde{\rho}_{xx} &\tilde{C}_{xxxx}q^2 -m\omega^2\rho_{n,xx} \\
  \end{bmatrix},\\
  M_t &=\left(\tilde{C}_{xyxy}+\frac{1}{4}A+\frac{1}{2}B\right)q^2  -m\omega^2\rho_{n,yy}.
\end{align}
The first block couples the (longitudinal)  $[\phi,u_x]^T$ components of the wave, and $M_t$ corresponds to $u_y$.
The speeds of sound follow from the same procedure developed in Sec.~\ref{Sec:Iso}. The coupled longitudinal speeds of sound $c_{\pm}(\hat{\mathbf{q}}_x)$ are determined from Eq.~(\ref{gencpm}), with
 \begin{align}
c_s^2 = \frac{\rho_{s,xx}\alpha_{\rho\rho}}{m},\quad c_l^2 = \frac{\tilde{C}_{xxxx}}{m\rho_{n,xx}},\quad 
g^2= \frac{\tilde{\rho}_{xx}^2\alpha_{\rho\rho}}{m\rho_{n,xx}},
\end{align}
while the transverse speed of sound is given by
\begin{align}
c_t(\hat{\mathbf{q}}_x)=\sqrt{\frac{\tilde{C}_{xyxy}+\frac{1}{4}A+\frac{1}{2}B}{m\rho_{n,yy}}}.\label{ctx}
\end{align}

\subsubsection{Propagation along the $y$-principal axis}
We now solve for the sound waves propagating along $\Gamma$-$Y$ with $\hat{\mathbf{q}}_l=\hat{\mathbf{q}}_y$, where  (\ref{Mgen}) simplifies to 
\begin{align} 
M_l&=\begin{bmatrix}\hbar^{2}\!\left(\!
\frac{\rho_{s,yy}q^{2}}{m}-\frac{\omega^{2}}{\alpha_{\rho\rho}}\!\right)   & 
-\hbar\omega q\tilde{\rho}_{yy} \\
-\hbar\omega q\tilde{\rho}_{yy} &\tilde{C}_{yyyyy}q^2 -m\omega^2\rho_{n,yy} \\
  \end{bmatrix},\\
  M_t &=\left(\tilde{C}_{xyxy}+\frac{1}{4}A-\frac{1}{2}B\right)q^2  -m\omega^2\rho_{n,xx}.
\end{align}
The first block couples the (longitudinal)  $[\phi,u_y]^T$ components of the wave, and $M_t$ corresponds to $u_x$. The speeds of sound in this case are $c_{\pm}(\hat{\mathbf{q}}_x)$ from Eq.~(\ref{gencpm}), with
 \begin{align}
c_s^2 = \frac{\rho_{s,yy}\alpha_{\rho\rho}}{m},\quad c_l^2 = \frac{\tilde{C}_{yyyy}}{m\rho_{n,yy}},\quad 
g^2= \frac{\tilde{\rho}_{yy}^2\alpha_{\rho\rho}}{m\rho_{n,yy}},
\end{align}
and
\begin{align}
c_t(\hat{\mathbf{q}}_y)=\sqrt{\frac{\tilde{C}_{xyxy}+\frac{1}{4}A-\frac{1}{2}B}{m\rho_{n,xx}}}.\label{cty}
\end{align}

\subsection{Stripe supersolid}\label{Sec:stripehydrosound}
For the stripe supersolid we can apply the simplifications discussed earlier, most importantly the displacement field reduces to $u_y$, and the non-trivial elastic parameters are $\{\rho_{n,yy},\alpha_{\rho\rho},C_{yyyy},\gamma_{yy},A\}$. As a result Eq.~(\ref{Mgen}) simplifies to 
\begin{align} 
M=
\begin{bmatrix}\hbar^2\!\left(\!
 \frac{ \rho_0 q_{x}^{2}}{m}+\frac{\rho_{s,yy} q_{y}^{2}}{m}-\frac{ \omega^{2}}{\alpha_{\rho\rho}}\!\right) \!\! 
& -\hbar\omega q_{y}\tilde{\rho}_{yy}\\
 -\hbar\omega q_{y}\tilde{\rho}_{yy} &\!\! Aq_x^2+\tilde{C}_{yyyy}q_y^2-m\omega^2\rho_{n,yy}
\end{bmatrix}\label{Mxstripe}
\end{align}
We  analyse propagation in the direction  $\hat{\mathbf{q}}_l=\cos\theta\hat{\mathbf{q}}_x+\sin\theta\hat{\mathbf{q}}_y$, parameterized by the planar angle $\theta$. This yields two coupled speeds of sound $c_{\pm}(\theta)$ given by Eq.~(\ref{gencpm}) with
\begin{align}
c_s^2(\theta) &= \frac{(\rho_0 \cos^2\theta+\rho_{s,yy}\sin^2\theta)\alpha_{\rho\rho}}{m},\\
 c_L^2(\theta) & =\frac{(A\cos^2\theta+\tilde{C}_{yyyy}\sin^2\theta)}{m\rho_{n,yy}},\\
g^2(\theta)&= \frac{\tilde{\rho}_{yy}^2\alpha_{\rho\rho}\sin^2\theta}{m\rho_{n,yy}}.
\end{align}
The matrix $M$ couples the  $[\phi,u_y]^T$ fields, with $\phi$ describing longitudinal superfluid motion, and $u_y$ being a longitudinal displacement for propagation along $\hat{\mathbf{q}}_y$  and a transverse displacement for propagation along $\hat{\mathbf{q}}_x$. 
Furthermore for propagation along $\hat{\mathbf{q}}_x$ the coupling term $g$ vanishes, and we have 
\begin{align}
c_+^2(0)=\frac{\rho_0\alpha_{\rho\rho}}{m},\qquad 
c_-^2(0)=\frac{A}{m\rho_{n,yy}},\label{cpm0stripe}
\end{align}
being the square speeds of sound for a longitudinal superfluid and a transverse stripe displacement wave, respectively. 
In contrast, for propagation along $\hat{\mathbf{q}}_y$ both sound branches are longitudinal and are coupled. For any other angle, the sound waves have mixed polarization character.

\section{Results} \label{Sec:Results}
Here we present results for the speeds of sound of anisotropic dipolar supersolids using calculations performed in experimentally relevant parameter regimes. These calculations also allow us to validate the hydrodynamic predictions for the speeds of sound  to those obtained directly from the full BdG calculations. 
To use the hydrodynamic theory we need to obtain the relevant elastic parameters (see Sec.~\ref{Sec:ElasticParams}). These can be calculated from a finite difference approximation to the second derivative of ground state energy density with respect to changes the hydrodynamic fields (also see Appendix \ref{Sec:EMFthry}). The speeds of sound are then predicted using the analytic results developed in Sec.~\ref{Sec:Sound}. From the BdG calculations we instead use the excitation energies of the lowest gapless bands ($\nu\le3$ for 2D crystalline states, and $\nu\le2$ bands for stripe state), at small values of $q$, and determine speeds of sound directly  as $c_{\nu}(\hat{\mathbf{q}})=\lim_{q\to0}\epsilon_{\nu\mathbf{q}}/q$. In practice the BdG results are difficult to compute accurately for $qa\lesssim10^{-3}$, but this is usually adequate to obtain the speeds of sound to a relative error of $\lesssim1\%$.
\subsection{Anisotropic triangular phase}\label{Sec:anisotriresults}
  \begin{figure}[htbp] 
    \centering
    \includegraphics[width=3.4in]{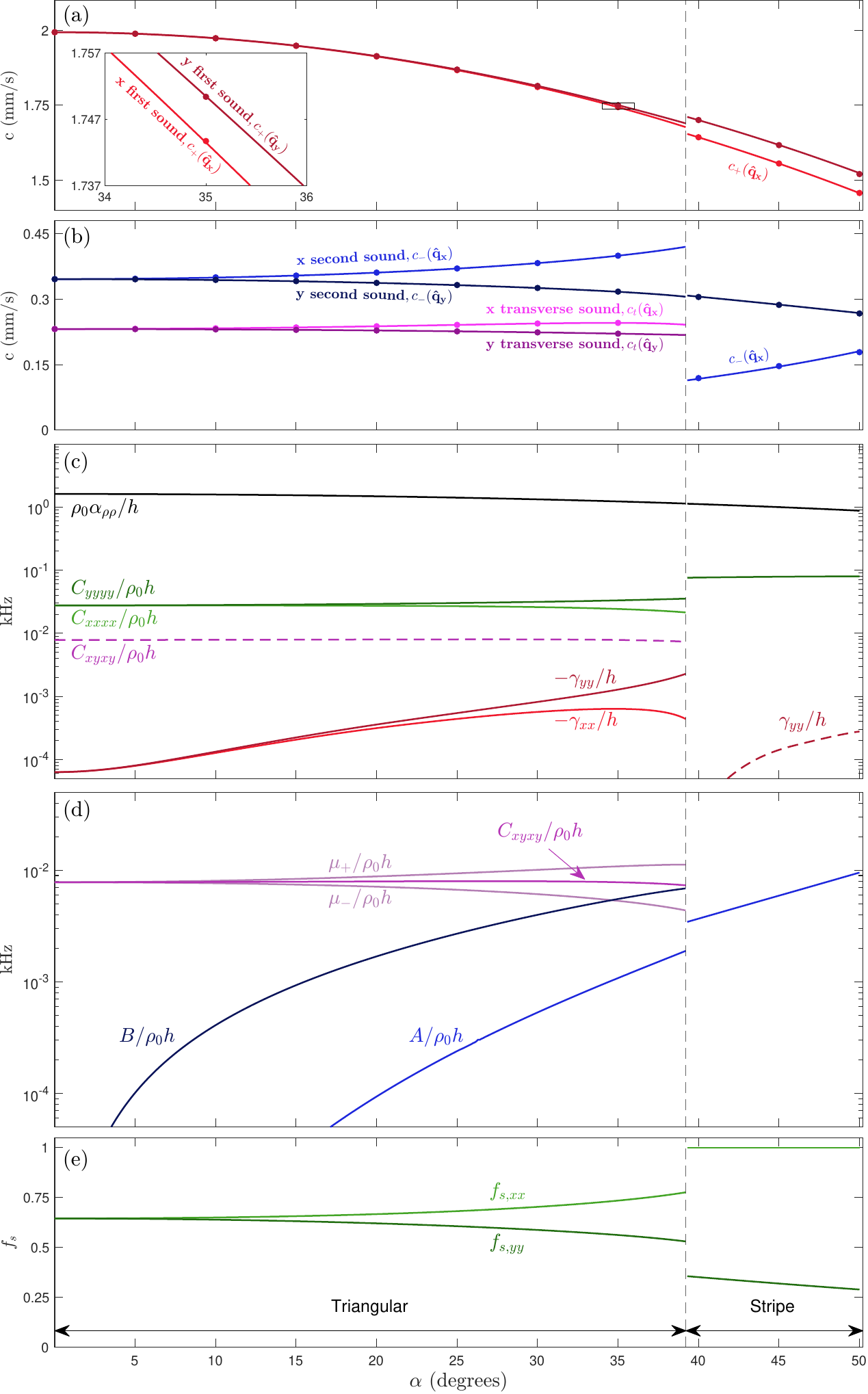} 
    \caption{Speeds of sound and elastic parameters for an anisotropic triangular supersolid. (a) First sound speeds ($c_+$) and (b) second ($c_-$) and transverse ($c_t$) sound speeds along principal axes as a function of dipole tilt. The markers indicate the speeds of sound obtained from BdG calculations whereas the lines are from the hydrodynamic theory. Inset to subplot (a) reveals the difference in speeds near $\alpha=35^\circ$. (c,d) Corresponding elastic parameters obtained from ground state calculations. Subplot (d) focuses on parameters relevant to the transverse speeds of sound. (e) Corresponding superfluid fractions obtained from ground state calculations. The vertical dashed line indicates the first-order transition of the ground state between triangular and stripe phases. Other parameters as in Fig.~\ref{fig:gsbs}. }
    \label{fig:tri}
 \end{figure}

Results for the elastic properties and speeds of sound of a dipolar supersolid are given in Fig.~\ref{fig:tri}. Here we consider the supersolid as a function of the dipole tilt $\alpha$, starting from the isotropic triangular state at $\alpha=0$, progressing through an anisotropic triangular supersolid, until a first order transition is crossed to a stripe state at $\alpha\approx39^\circ$. In this subsection we focus on the triangular state, and discuss the stripe state properties in Sec.~\ref{Sec:striperesults}.

In Fig.~\ref{fig:tri}(a) and (b) we examine the speeds of sound for propagation along the $x$- and $y$-principal axes  [cf.~Fig.~\ref{fig:gsbs}(c)]. The hydrodynamic results are seen to be in excellent agreement with those of the BdG calculation.
For the triangular state, the transverse excitation branch $c_t$ has the slowest sound speed, with the lower longitudinal branch $c_-$ lying slightly above it, while the upper longitudinal branch $c_+$ is much faster than the other two. The sound velocities are anisotropic, differing along the two principal axes for the same excitation branch. This anisotropy is most pronounced for the lower longitudinal branch ($c_-$) and the transverse branch ($c_t$), for which the sound speeds along $\hat{\mathbf{q}}_x$ exceed those along $\hat{\mathbf{q}}_y$, with the difference increasing as $\alpha$ grows. In contrast, for the upper longitudinal branch the sound speed along $\hat{\mathbf{q}}_y$ is larger than that along $\hat{\mathbf{q}}_x$, although the relative difference is smaller than for the lower two branches.

In Figs.~\ref{fig:tri}(c)-(e) we show the relevant elastic parameters used in the hydrodynamic theory. As observed in earlier work, dipolar supersolids are dominated by the bulk modulus $\alpha_{\rho\rho}$, which is significantly larger than the lattice elastic parameters.  The effect of dipole tilt in creating anisotropy is revealed in the elastic parameters. At $\alpha=0$ the elastic parameters are isotropic (see Sec.~\ref{Sec:Iso}), with  $C_{xxxx}=C_{yyyy}=\lambda+2\mu$, $C_{xyxy}=\mu$, $C_{xxyy}=\lambda$ (not shown in Fig.~\ref{fig:tri}), also $\gamma_{xx}=\gamma_{yy}$ and $f_{s,xx}=f_{s,yy}$.
For the isotropic case the orientational parameters $\{A,B\}$ are zero.
As $\alpha$ increases the parameter pairs distinguishing the response along $x$ and $y$, i.e., $\{C_{xxxx},C_{yyyy}\}$,   $\{\gamma_{xx},\gamma_{yy}\}$,  and $\{f_{s,xx},f_{s,yy}\}$ become increasingly different. We also observe that the orientational parameters $\{A,B\}$  increase rapidly with $\alpha$. In subplot (d) we show the effective shear moduli  $\mu_\pm\equiv C_{xyxy}+(A\pm2B)/4$, which combine the shear modulus $C_{xyxy}$ with the orientational parameters. The effective shear modulus $\mu_+$ determines the $q_x$ transverse speed of sound   (\ref{ctx}), while $\mu_-$ determines the $q_y$ transverse speed of sound  (\ref{cty}).  These results demonstrate the important role of the orientational terms on the transverse sound speeds.
  
\subsection{Stripe phase}  \label{Sec:striperesults}
Here we present results for the stripe phase and test the hydrodynamic results for the speeds of sound developed in Sec.~\ref{Sec:stripehydrosound}. The first order transition to the stripe state is apparent in  Fig.~\ref{fig:tri}, occurring at a tilt angle of $\alpha\approx39^\circ$. The results in this figure also emphasize the reduction in the number of elastic parameters needed for the stripe state relative to the 2D crystalline state. This reflects the single broken translational symmetry, as discussed in Sec.~\ref{Sec:HydroFormalism}.

\begin{figure}[htbp] 
    \centering
    \includegraphics[width=3.4in]{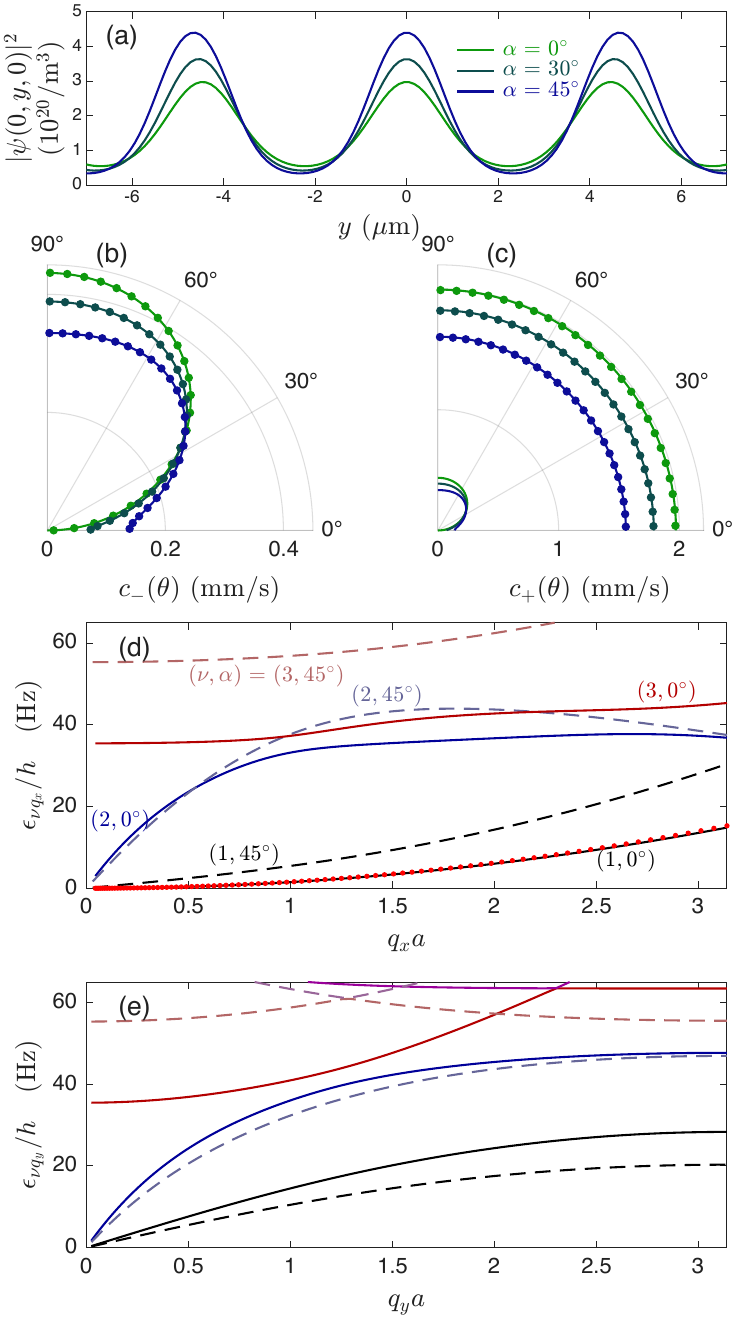} 
    \caption{Stripe phase density profile and excitation properties. (a) Density of the stripe state on the $y$-axis for the tilt angles $\alpha=0^\circ,30^\circ,$ and $45^\circ$, as labeled in the subplot. Speeds of sound versus propagation angle $\theta$ for the (b) lower $c_-$ and (c) upper $c_+$ branch. Solid lines are hydrodynamic results and markers indicate speeds of sound obtained from BdG calculations. Hydrodynamic $c_-$ results   are shown in (c) for context. The lowest three excitations bands ($\nu=1,2,3$) for propagation along (d) ${q}_x$ and (e) ${q}_y$ obtained from BdG calculations. Results shown for $\alpha=0^\circ$ (solid) and  $\alpha=45^\circ$ (dashed).  Other parameters for this calculation are given in Table \ref{tab:stripeparams}.}
    \label{fig:stripes}
 \end{figure}

Our main results for the stripe phase are presented in Fig.~\ref{fig:stripes}, revealing the speeds of sound and band structure. Here results are given for several tilt angles, and the corresponding elastic parameters are given in Table \ref{tab:stripeparams}.
In Fig.~\ref{fig:stripes}(a) we show the stripe state density on the $y$-axis. As the tilt angle increases, the peak density, the contrast of the density modulation and the lattice constant, all increase. This correlates with a decrease in the bulk modulus and an increase in the elastic modulus (see Table \ref{tab:stripeparams}). In subplots (b) and (c) the $c_\pm$ speeds of sound are shown as a function of propagation direction. The $c_-$ speed of sound is strongly anisotropic, and vanishes for propagation along ${q}_x$ when $\alpha=0$  [where $A=0$, also see Eq.~(\ref{cpm0stripe})]. In contrast,  $c_+$ is almost independent of propagation direction. For both branches the hydrodynamic predictions for the speeds of sound are in good agreement with the BdG results. 

Figures \ref{fig:stripes}(d) and (e) show the excitations spectra  along ${q}_x$  and  ${q}_y$, respectively, for two values of dipole tilt. 
For the untilted results in Fig.~\ref{fig:stripes}(d), the lowest band is quadratic, exhibiting a free particle dispersion relation. This is  consistent with $c_-$ vanishing for $\hat{\mathbf{q}}_x$ propagation. For non-zero tilt, the lowest branch becomes linear in $q_x$. In contrast the both gapless branches are linear in $q_y$ at low momentum in Fig.~\ref{fig:stripes}(e) irrespective of the tilt angle.

 \begin{table}[htbp]
    \centering 
    \begin{tabular}{@{} lllll@{}}  
       \toprule  
     $\alpha$ & & $0^\circ\qquad$  & $30^\circ\qquad$ &$45^\circ\qquad$\\
$a$ & ($\mu$m) & 4.46 & 4.54 & 4.65\\
       \midrule  
$f_{s,yy}$ &  & 0.70 & 0.57 & 0.45\\
$\rho_0\alpha_{\rho\rho}/h$ & (kHz) & 1.60 & 1.31 & 1.00\\
$C_{yyyy}/\rho_0h$ & (Hz) & 33.6 & 48.7 & 60.0\\
$A/\rho_0h$ & (Hz) & 0.00 & 0.93 & 4.39\\
$\gamma_{yy}/h$ & (Hz) & -1.26 & -0.89 & 0.21\\
       \bottomrule
    \end{tabular}
    \caption{Elastic parameters for stripe state calculations of a $^{164}$Dy system with $\omega_z/2\pi=72.4\,$Hz, $\rho_0=0.075/a_{dd}^2$ and $a_s= 101.4\,a_0$.}
    \label{tab:stripeparams}
 \end{table}

\begin{figure}[htbp] 
\centering
 \includegraphics[width=3.4in]{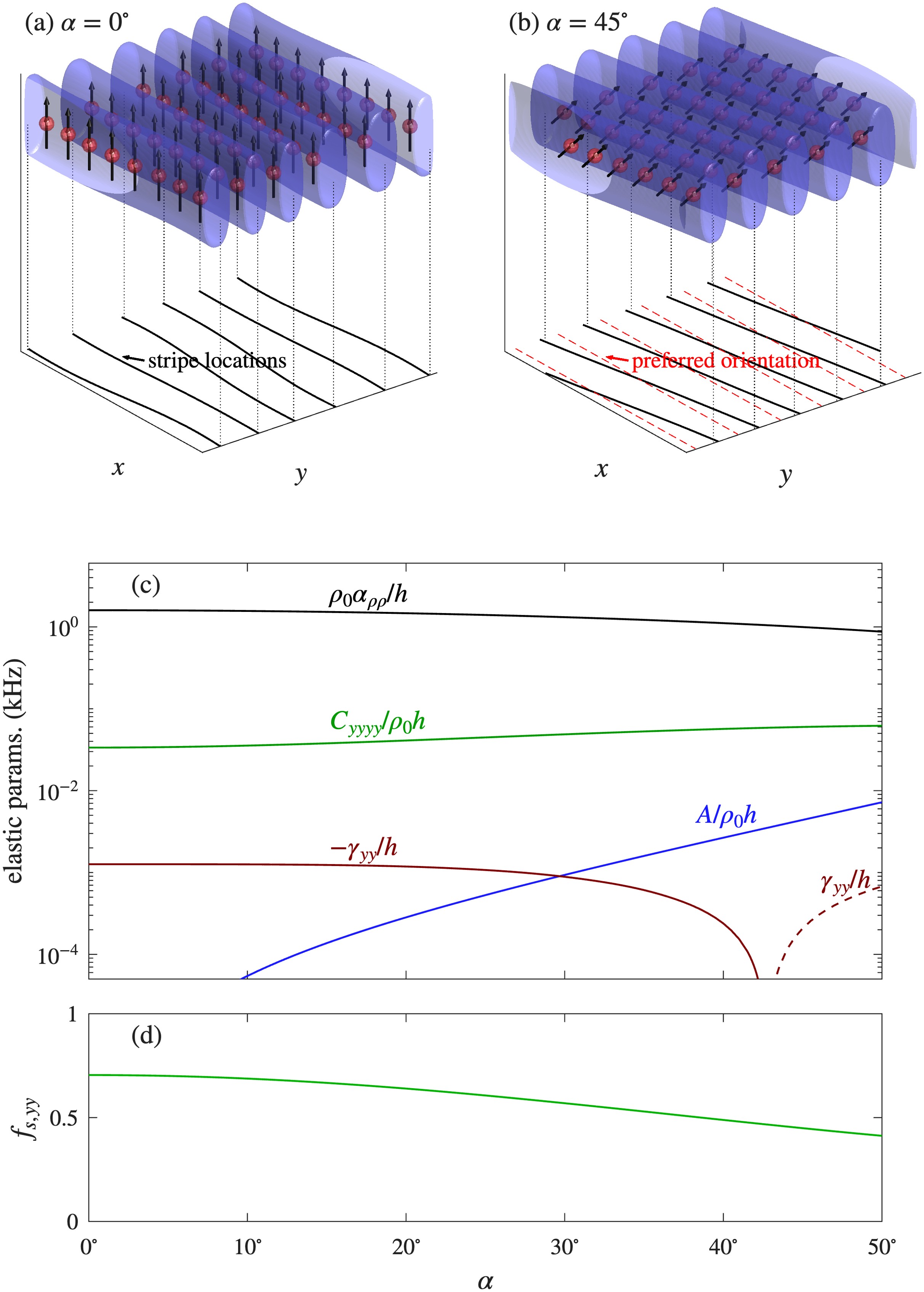} 
 \caption{Superfluid smectic interpretation of the stripe state. Schematic of smectic stripe state with (a) untilted and (b) tilted dipoles. For tilted dipoles the smectic stripe layers prefer to align parallel to the $x$ axis. (c), (d) The elastic parameters for the stripe state as a function of tilt angle. Other system parameters in Table \ref{tab:stripeparams}.  }
\label{fig:SFsmetic}
\end{figure}
 
\subsubsection{2D smectic material interpretation}
 In the smectic phase of a liquid crystal, the constituent molecules are arranged in layers \cite{deGennesProst1993}.
 Within each layer, the molecules can move around like in a liquid, but motion between layers is restricted. The smectic phase thus has a broken translational symmetry along one direction (perpendicular to the layers) and a fluid-like behavior within the layers. The qualitative similarity of a smectic material to the stripe-crystalline structure motivates it as an alternative description of a stripe supersolid \cite{Hofmann2021a} (also see \cite{Sindik2024a,Zawislak2025a}).  In a smectic system the elastic contribution of the layers is described by a term analogous to our formulation, i.e.,~as $\frac{1}{2}C_{yyyy}(\partial_yu_y)^2$, with  $C_{yyyy}$ being referred to as the layer compressibility and $u_y$ is related to a quantity known as the layer phase  \cite{Chaikin1995a} (see Fig.~\ref{fig:SFsmetic}). Our formalism in this manuscript immediately connects to such interpretation, but emphasises the necessity of adding the orientational term  $\frac{1}{2}A(\partial_xu_y)^2$ for the description of dipole tilt. This term has not been included in previous treatments based on the smectic analogy.
 In Fig.~\ref{fig:SFsmetic}(c), (d) we examine the dependence of the elastic parameters as a function of dipole tilt angle. This quantifies the typical behavior of the elastic parameters, and shows how they can be controlled by dipole tilt.
 
\subsubsection{Relation to stripe BEC}
Recent experiments \cite{Chauveau2023a,Tao2023a} produced stripe BECs with the application of an optical lattice. Unlike supersolids, the translational symmetry in this system is not broken spontaneously and the system only has a single sound branch. By measuring the speeds of sound propagating parallel and perpendicular to the stripes, the experiments verified the Leggett prediction for the reduction of the superfluid fraction in a supersolid \cite{Leggett1970a,Leggett1998a}.  

We can specialize our theory to this case. We take the optical lattice vector to be along $y$. The system density is thus uniform along $x$ and $\rho_{s,xx}=\rho_0$, while  $\rho_{s,yy}<\rho_0$ due to the imposed density modulation. As the optical lattice is rigid, $C_{yyyy}$ and $A$ can be taken as large, so that all dynamics in $u_y$ is frozen out, and the system only has a single speed of sound. This is described by Eq.~(\ref{EL1}) specialized to this case:
\begin{align}
{\hbar^{2}}\left(\frac{1}{\alpha_{\rho\rho}}\partial_{t}^{2}-\frac{\rho_{s,ij}}{m}\partial_{ij}\right)\phi=0.
\end{align}
Solving this, the speed of sound along the principal axes are  
\begin{align}
c(\hat{\mathbf{q}}_x)&=\sqrt{\frac{\rho_0\alpha_{\rho\rho}}{m}},\\
c(\hat{\mathbf{q}}_y)&=\sqrt{\frac{\rho_{s,yy}\alpha_{\rho\rho}}{m}}.
\end{align}
This reveals how a measurement of the sound speed can be used to determine the superfluid fraction. 
In contrast, the relationship between the speeds of sound and superfluidity is more involved for a stripe supersolid [see Sec.~\ref{Sec:stripehydrosound}]. A method to determine the superfluid fraction for the 1D dipolar supersolid has been discussed in Ref.~\cite{Sindik2024a}, and is applicable to stripe supersolids. We also note the development of a Josephson oscillation technique for determining the superfluid fraction of a 1D dipolar supersolid \cite{Biagioni2024a,Platt2025a}

\section{Outlook and Conclusions}\label{Sec:concl}
In this work we have presented a theory for the excitations of anisotropic planar supersolids.  We developed a hydrodynamic framework for describing the long-wavelength excitations of anisotropic dipolar supersolids, and validated it against full BdG calculations for triangular and stripe phases. By identifying the relevant elastic parameters -- including two orientational coefficients that quantify the broken rotational symmetry produced by dipole tilt -- we are able to describe how anisotropy enters the supersolid excitation spectrum. The analytic expressions we obtained for the sound speeds reveal the role of orientational elasticity, particularly in strongly tilted or stripe-like configurations. Our results provide a basis for interpreting collective-mode measurements in dipolar supersolids and offer a route toward characterizing anisotropic supersolid order in emerging experimental platforms, including systems that realize stripe phases via optical lattices or other engineered modulations.

There are a number of avenues for future work in this area. First, to extend the ideas to finite planar systems, either in pancake shaped harmonic traps or a 2D box trap (e.g.,~see \cite{Juhasz2022a,He2025a,Zhen2025a}). Here the hydrodynamic properties would manifest the low-energy collective modes which could be measured by excitation spectroscopy. Another direction would be to consider stripe supersolids produced in spin-orbit coupled systems \cite{Li2017a,Putra2020a,Chisholm2024a}. This system has many similarities to the dipolar stripe state we have consider, however as the system is not Galilean invariant, the hydrodynamic theory we have developed does not directly apply to the spin-orbit case.

\textit{Note added.} During the late stages of preparation, we became aware of the experimental production of a stripe supersolid using an Er BEC in a quasi-2D trap by tilting the dipoles partially into the plane of the trap and reducing the $s$-wave scattering length \cite{He2025b}.

\section*{Acknowledgments} 
\appendix

The authors acknowledge useful conversations with D.~Baillie, B.~Ripley, and L.~Chomaz. They also acknowledge high-performance computing resources from the New Zealand eScience Infrastructure (NeSI) and funding from the Marsden Fund of the Royal Society of New Zealand.

\section{Extended mean-field theory}\label{Sec:EMFthry}
In this appendix we outline the extended mean-field theory used to calculate stationary supersolid states and their excitations. These types of calculations have been used previously for 1D and 2D supersolid cases in Refs.~\cite{Blakie2023a,Platt2024a,Poli2024b,Blakie2025a}. Here the main difference from the previous 2D supersolid calculations is the tilt of the dipoles from the $z$-axis.    
In general, our calculations are performed on a single unit cell (uc).
\paragraph{2D supersolid}   the uc is given by a primitive unit cell in the $xy$-plane spanned by the lattice vectors $\{\mathbf{a}_\pm\}$ [see Eq.~(\ref{apm})], and with $z\in(-\infty,\infty)$. Here the wave functions (ground state and excitations) satisfies twisted-phase periodic boundary conditions in the plane, i.e.,~$f(\bm{\rho}+\mathbf{a}_\beta,z)=f(\bm{\rho},z)e^{i\mathbf{q}\cdot\mathbf{a}_\beta}$, where $\hbar\mathbf{q}$ is a planar quasimomentum vector\footnote{For analysis of the superfluid fraction, the superfluid velocity is proportional to the quasimomentum. }.  \paragraph{Stripe supersolid} for our choice of geometry, the stripe wave function is uniform along $x$, and can be reduced to the 2D field $\psi(y,z)$. The unit cell along $y$ is defined by $a$, and the ground and excited state have boundary conditions $f(y+a,z)=f(y,z)e^{iq_ya}$. 
 
 For both types of supersolid states we discretize the numerical solutions on the unit cell using a finite interval of length $L_z$ along $z$. This allows us to compute kinetic energy operators and density convolutions with spectral accuracy using Fourier spectral methods. For the dipole interaction, we make use of the $z$-cutoff dipole-dipole interaction potential to correct for the finite $L_z$ effect on long-range interactions (see \cite{Ronen2006a}). 

\subsection{Extended Gross-Pitaevskii theory for stationary states}
The energy density function for a dipolar BEC in a planar trap is\footnote{We hereon describe the method for 2D supersolids. The adaption to the stripes is immediate. E.g., in Eq.~(\ref{Eden}) we set $\bx\to (y,z)$ and $A_{uc}\to a$.}
\begin{align} 
\mathcal{E}&\equiv \frac{1}{A_{\mathrm{uc}}} \int_{\mathrm{uc}} \!d\mathbf{x}\, \Psi_0^*\left(H_\mathrm{sp}+\Phi(\bx)  +\gammaQF|\Psi_0(\bx)|^3\right)\Psi_0,\label{Eden}
\end{align}
where 
\begin{align}
H_\mathrm{sp}&=-\frac{\hbar^2}{2m}\nabla^2 + \frac12 m\omega_z^2z^2,\\
\Phi(\bx)&=\int d\bx'\,U(\bx-\bx')|\Psi_0(\bx')|^2.
\end{align}
Here the interaction potential is
\begin{equation}
	U(\mathbf{r}) = \frac{4\pi a_s\hbar^2}{m}\delta(\br) + \frac{3\add\hbar^2}{m r^3}\left(1-3\frac{(\hat{\mathbf{e}}\cdot{\mathbf{r}})^2}{r^2}\right),
\end{equation}
where $ \textbf{r}= \textbf{x}- \textbf{x}'$ is the relative separation between the particles and $\hat{\mathbf{e}}=\cos\alpha\hat{\mathbf{z}}+\sin\alpha\hat{\mathbf{x}}$ is the polarization direction of the dipoles. Here $a_s$ is the $s$-wave scattering length,  $\add = m\mu_0\mu_m^2/12\pi\hbar^2$ is the dipole length, and $\mu_m$  is the atomic magnetic moment. The effects of quantum fluctuations are described by the term with coefficient $\gammaQF = \frac{128\pi\hbar^2}{3m}a_s\sqrt{\frac{a_s^3}{\pi}}\Re\left\{\mathcal{Q}_5(\add/a_s)\right\}$, where  \cite{Lima2011a}
\begin{align}
\mathcal{Q}_5(x)=\int_0^1 du[1+x(3u^2 - 1)]^{5/2}.
\end{align}

Numerically, we find solutions using the gradient flow method \cite{Bao2010a,Lee2021a}. This determines the minimum energy solution on a specified unit cell. In this process we also adjust the lattice vectors  to find the values $\{\mathbf{a}_j\}$ that minimise the energy density. The resulting stationary state is thus specified by the interaction and trap parameters, and for  $\{\rho,\mathbf{a}_j,\mathbf{v}=\mathbf{0}\}$. Elastic parameters are then calculated by making slight variations in this set of parameters recalculating the $\mathbf{E}$ under those conditions, and then evaluating the elastic parameters by finite differences  (see Sec.~\ref{Sec:ElasticParams}).

Stationary state solutions  $\Psi_0(\mathbf{x})$ for the condensate wave function can be taken as real, and satisfies the time-independent extended Gross-Pitaevskii equation  $\mathcal{L}\Psi_0=\mu\Psi_0$, where
\begin{align} 
\mathcal{L}&\equiv H_\mathbf{sp} + \Phi   +\gammaQF|\Psi_0|^3,
\end{align}
and $\mu$ is the chemical potential. 
The constraint that the system has an average areal density of $\rho_0$ is enforced by the normalization condition $\int_{\mathrm{uc}}d\mathbf{x}|\Psi_0(\mathbf{x})|^2=\rho_0 A_{\mathrm{uc}}$.

\subsection{Extended Bogoliubov-de Gennes theory for excitations}
The equations for the quasi-particle excitations of the supersolid can be obtained by linearizing the time-dependent extended Gross-Pitaevskii equation, $i\hbar\dot{\Psi}=\mathcal{L}\Psi$ about a ground state as
\begin{align}
\Psi(\bx,t)=& e^{-i\mu t/\hbar}\biggl[\Psi_0(\bx)+\sum_{\nu\mathbf{q}}\left\{ c_{\nu\mathbf{q}} u_{\nu\mathbf{q}}(\bx)e^{-i\epsilon_{\nu\mathbf{q}}t/\hbar}\right.   \nonumber\\
& \left.   -c^*_{\nu\mathbf{q}} v^*_{\nu\mathbf{q}}(\bx) e^{i\epsilon^*_{\nu\mathbf{q}}t/\hbar} \right\} \biggl],
\end{align}
where $\{ c_{\nu\mathbf{q}}\}$ are the expansion amplitudes, and $\nu$ and $\mathbf{q}$ are the band index and planar quasimomentum of the excitation, respectively. Here $\{u_{\nu \mathbf{q}}(\mathbf{x}),v_{\nu \mathbf{q}}(\mathbf{x})\}$ are the quasiparticle amplitudes, with respective energies $\{\epsilon_{\nu\mathbf{q}}\}$. These are determined by solving the BdG equations 
\begin{align}
\begin{bmatrix} 
      \mathcal{L}+X-\mu & -X \\
      X & -(\mathcal{L}+X-\mu) \\
   \end{bmatrix} \begin{bmatrix} u_{\nu\mathbf{q}} \\ v_{\nu\mathbf{q}}  \end{bmatrix} =\epsilon_{\nu\mathbf{q}}\begin{bmatrix} u_{\nu\mathbf{q}} \\ v_{\nu\mathbf{q}}  \end{bmatrix}.
\end{align} 
The functions $u_{\nu\bq}$ and $v_{\nu\bq}$ have the Bloch form, i.e.,~$u_{\nu\bq}(\bx)=\bar{u}_{\nu\bq}(\bx)e^{i\bq\cdot\bx}$ with $\bar{u}_{\nu\bq}(\bx)$ periodic in the unit cell, and the exchange operator $X$ is defined so that 
\begin{align} 
\!Xf&=\!\Psi_0 \!\int\!d\mathbf{x}^\prime U(\mathbf{x}\!-\!\mathbf{x}^\prime)f(\mathbf{x}^\prime)\Psi_0(\mathbf{x}^\prime)  +\frac32\gamma_{\mathrm{QF}}|\Psi_0|^3f. 
\end{align}

  \begin{widetext}
 \section{Full Euler-Lagrange equations}\label{Sec:AppendEL}
 For the general 2D supersolid the Euler-Lagrange equations are:
 \begin{align}
\frac{\hbar^2}{\alpha_{\rho\rho}}\partial_{t}^2\phi-\frac{\hbar^2}{m}(\rho_{s,xx}\partial_{x}^2+\rho_{s,yy}\partial_{y}^2)\phi= \hbar\partial_t(\tilde{\rho}_{xx} u_{xx}+ \tilde{\rho}_{yy}u_{yy}),\hspace{7.2cm}\\
 m\rho_{n,xx}\partial_{t}^2u_x 
 -\partial_x(\tilde C_{xxxx}u_{xx}+\tilde C_{xxyy}u_{yy}) -\partial_y[\tilde C_{xyxy}(u_{xy}+ u_{yx})+\tfrac{1}{4}A(u_{xy}- u_{yx})-\tfrac{1}{2}Bu_{xy}]=\hbar\tilde{\rho}_{xx} \partial_t\partial_x\phi,\\
 m\rho_{n,yy}\partial_{t}^2u_y 
 -\partial_y(\tilde C_{yyyy}u_{yy}+\tilde C_{xxyy}u_{xx})-\partial_x[\tilde C_{xyxy}(u_{xy}+ u_{yx})-\tfrac{1}{4}A(u_{xy}- u_{yx})+\tfrac{1}{2}Bu_{yx}]= \hbar \tilde{\rho}_{yy}\partial_t\partial_y\phi.
 \end{align}
For the stripe supersolid they are
  \begin{align}
\left[\frac{\hbar^2}{\alpha_{\rho\rho}}\partial_{t}^2-\frac{\hbar^2}{m}(\rho_0\partial_{x}^2+\rho_{s,yy}\partial_{y}^2)\right]\phi&=  \hbar\tilde{\rho}_{yy}\partial_tu_{yy},\\ 
 (m\rho_{n,yy}\partial_{t}^2 
 -\tilde C_{yyyy}\partial_y^2  -A\partial_x^2)u_y&= \hbar \tilde{\rho}_{yy}\partial_t\partial_y\phi.
 \end{align}
 \end{widetext}

%

\end{document}